\documentclass[aps,superscriptaddress,onecolumn,preprint,citeautoscript,]{revtex4-1}
\usepackage{amssymb,amsfonts,amsmath,bm}
\usepackage{graphicx}
\setcitestyle{super}

\begin{document}
    \makeatletter
    \renewcommand*{\@fnsymbol}[1]{\ensuremath{\ifcase#1\or *\or *\or *\or
       \mathsection\or \mathparagraph\or \|\or **\or \dagger\dagger
       \or \ddagger\ddagger \else\@ctrerr\fi}}
    \makeatother

\title{Nanoscale NMR Spectroscopy and Imaging of Multiple Nuclear Species}

\author{Stephen J. DeVience}
\email{devience@fas.harvard.edu}
\thanks{Corresponding author}
\affiliation{Department of Chemistry and Chemical Biology, Harvard University, 12 Oxford St., Cambridge, MA 02138, USA.}

\author{Linh M. Pham}
\affiliation{Harvard-Smithsonian Center for Astrophysics, 60 Garden St., Cambridge, MA 02138, USA.}

\author{Igor Lovchinsky}
\affiliation{Department of Physics, Harvard University, 17 Oxford St., Cambridge, MA 02138, USA.}

\author{Alexander O. Sushkov}
\affiliation{Department of Chemistry and Chemical Biology, Harvard University, 12 Oxford St., Cambridge, MA 02138, USA.}
\affiliation{Department of Physics, Harvard University, 17 Oxford St., Cambridge, MA 02138, USA.}

\author{Nir Bar-Gill}
\affiliation{Department of Applied Physics and Racah Institute of Physics, Hebrew University, Edmond J. Safra campus, Jerusalem, Israel.}

\author{Chinmay Belthangady}
\affiliation{Harvard-Smithsonian Center for Astrophysics, 60 Garden St., Cambridge, MA 02138, USA.}

\author{Francesco Casola}
\affiliation{Harvard-Smithsonian Center for Astrophysics, 60 Garden St., Cambridge, MA 02138, USA.}

\author{Madeleine Corbett}
\affiliation{School of Engineering and Applied Sciences, Harvard University, 15 Oxford St., Cambridge, MA 02138, USA.}

\author{Huiliang Zhang}
\affiliation{Department of Physics, Harvard University, 17 Oxford St., Cambridge, MA 02138, USA.}

\author{Mikhail Lukin}
\affiliation{Department of Physics, Harvard University, 17 Oxford St., Cambridge, MA 02138, USA.}

\author{Hongkun Park}
\affiliation{Department of Chemistry and Chemical Biology, Harvard University, 12 Oxford St., Cambridge, MA 02138, USA.}
\affiliation{Department of Physics, Harvard University, 17 Oxford St., Cambridge, MA 02138, USA.}
\affiliation{Center for Brain Science, Harvard University, 52 Oxford St., Cambridge, MA 02138, USA.}

\author{Amir Yacoby}
\affiliation{Department of Physics, Harvard University, 17 Oxford St., Cambridge, MA 02138, USA.}
\affiliation{School of Engineering and Applied Sciences, Harvard University, 15 Oxford St., Cambridge, MA 02138, USA.}

\author{Ronald L. Walsworth}
\email{rwalsworth@cfa.harvard.edu}
\thanks{Corresponding author}
\affiliation{Harvard-Smithsonian Center for Astrophysics, 60 Garden St., Cambridge, MA 02138, USA.}
\affiliation{Department of Physics, Harvard University, 17 Oxford St., Cambridge, MA 02138, USA.}
\affiliation{Center for Brain Science, Harvard University, 52 Oxford St., Cambridge, MA 02138, USA.}

\maketitle

\begin{center}
{\bf \large{Abstract}}\\
\end{center}

Nuclear magnetic resonance (NMR) and magnetic resonance imaging (MRI) are well-established techniques that provide valuable information in a diverse set of disciplines but are currently limited to macroscopic sample volumes. Here we demonstrate nanoscale NMR spectroscopy and imaging under ambient conditions of samples containing multiple nuclear species, using nitrogen-vacancy (NV) colour centres in diamond as sensors. With single, shallow NV centres in a diamond chip and samples placed on the diamond surface, we perform NMR spectroscopy and one-dimensional MRI on few-nanometre-sized samples containing $^1$H and $^{19}$F nuclei. Alternatively, we employ a high-density NV layer near the surface of a diamond chip to demonstrate wide-field optical NMR spectroscopy of nanoscale samples containing $^1$H, $^{19}$F, and $^{31}$P nuclei, as well as multi-species two-dimensional optical MRI with sub-micron resolution. For all diamond samples exposed to air, we identify a ubiquitous $^1$H NMR signal, consistent with a $\sim 1$ nm layer of adsorbed hydrocarbons or water on the diamond surface and below any sample placed on the diamond. This work lays the foundation for nanoscale NMR and MRI applications such as studies of single proteins and functional biological imaging with subcellular resolution, as well as characterization of thin films with sub-nanometre resolution.
\newpage

Nuclear magnetic resonance (NMR) spectroscopy and magnetic resonance imaging (MRI) provide non-invasive information about multiple nuclear species in bulk matter, with wide-ranging applications from basic physics and chemistry to biomedical imaging \cite{Mansfield2}. However, the spatial resolution of conventional NMR and MRI is limited to several microns even at large magnetic fields ($> 1$ tesla) \cite{Glover1}, which is inadequate for many frontier scientific applications such as single molecule NMR spectroscopy and {\em in vivo} MRI of individual biological cells. A promising approach for nanoscale NMR and MRI exploits optical measurements of nitrogen-vacancy (NV) colour centres in diamond, which provide a combination of magnetic field sensitivity and nanoscale spatial resolution unmatched by any existing technology, while operating under ambient conditions in a robust, solid-state system \cite{TaylorNatPhys2008, MazeNature2008, BalasubramanianNature2008}.  Recently, single, shallow NV centres were used to demonstrate NMR of nanoscale ensembles of proton spins, consisting of a statistical polarization equivalent to ~100 - 1,000 spins in uniform samples covering the  surface of a bulk diamond chip \cite{Rugar1, Wrachtrup1}. Here, we realize nanoscale NMR spectroscopy and MRI of multiple nuclear species ($^{1}$H, $^{19}$F, $^{31}$P) in non-uniform (spatially-structured) samples under ambient conditions and at moderate magnetic fields ($\sim 20$ millitesla) using two complementary sensor modalities. We interrogate single shallow NV centres in a diamond chip to perform simultaneous multi-species NMR spectroscopy and one-dimensional MRI on few-nanometre-sized samples placed on the diamond surface, which have a statistical spin polarization equivalent to $\sim 100$ polarized nuclei. We also employ a diamond chip containing a shallow, high-density NV layer to demonstrate wide-field optical NMR spectroscopy and two-dimensional MRI with sub-micron resolution of samples containing multiple nuclear species. This work lays the foundation for diverse NMR and MRI applications at the nanoscale, such as determination of the structure and dynamics of single proteins and other biomolecules, identification of transition states in surface chemical reactions, and functional biological imaging with subcellular resolution and cellular circuit field-of-view.

The spatial resolution of conventional NMR and MRI is limited to macroscopic length scales due to the modest signal-to-noise ratio (SNR) provided by inductively-detected thermal spin polarization, even in large ($> 1$ tesla) magnetic fields, and the finite strength of externally-applied magnetic field gradients used for Fourier k-space imaging \cite{Glover1}. Other precision magnetic sensors have only macroscopic resolution, e.g., semiconductor Hall effect sensors \cite{Bending1} and atomic magnetometers \cite{Budker1}, and/or require operation at cryogenic temperatures or in vacuum, e.g., superconducting quantum interference devices (SQUIDs) \cite{Nowack2013imaging} and magnetic resonance force microscopy (MRFM) \cite{RugarNature2004,degen2009nanoscale}.  Alternatively, NV centres in room-temperature diamond can be brought within a few nanometres of magnetic field sources of interest while maintaining long NV electronic spin coherence times ($\sim 100$ $\mu$s), a large Zeeman shift of the NV spin states ($\sim 28$ MHz/mT), and optical preparation and readout of the NV spin (Fig.\ 1a).  Highlights of NV-diamond magnetic sensing to date, all performed under ambient conditions, include sensitive spectroscopy \cite{ChildressScience2006, BarGill2, belthangady2013dressed} and imaging \cite{kolkowitz2012sensing, taminiau2012detection, grinolds2014mri} of electron and nuclear spin impurities within the diamond sample; single electron spin imaging external to the diamond sensor \cite{grinolds2013nanoscale}; sensing the aforementioned nanoscale NMR of proton spins in samples placed on the diamond surface \cite{Rugar1, Wrachtrup1, Degen1}; targeted detection of single paramagnetic molecules attached to the diamond surface \cite{Sushkov1}; and wide-field magnetic imaging of living magnetotactic bacteria, with sub-micron resolution \cite{LeSageNature2013}.

In the first NV diamond sensor modality used in the present work (Fig.\ 1b), a scanning confocal microscope interrogates a single NV centre located a few nanometres below the surface of a high-purity diamond chip. In the second sensor modality (Fig.\ 1c), the fluorescence from a shallow (5 - 15 nm deep), high-density ($3.5 \times 10^{11}$ cm$^{-2}$) NV ensemble layer near the surface of a diamond chip is imaged onto a CCD camera \cite{Pham1}. The NV ensemble wide-field microscope provides pixel-by-pixel multi-species NMR spectroscopy and two-dimensional MRI with sub-micron resolution and wide field-of-view, in a robust device that does not rely on identifying and addressing an optimally chosen NV centre; while the single NV confocal microscope can extract thickness information of layered thin films containing different nuclear species, with sub-nanometer resolution.

\begin{center}
{\bf \large{The NV NMR Experiment}}\\
\end{center}

\begin{figure*}[t!]
\includegraphics[width=6.5in]{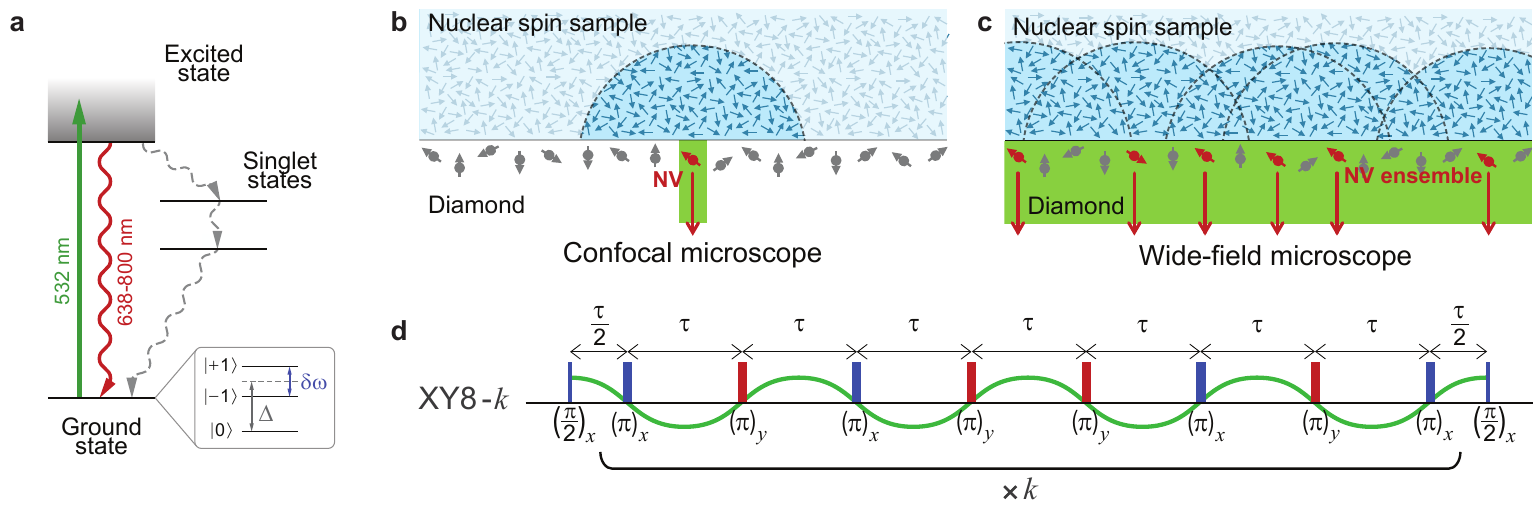}  
\caption{{\bf NV NMR Experiment.} {\bf a}, NV centre energy level diagram; see Methods for details. {\bf b}, A confocal microscope interrogates a single shallow NV centre, which detects NMR signals from a few-nanometre region of sample on the diamond surface. {\bf c}, A wide-field microscope images fluorescence from a shallow, high-density layer of NV centres allowing detection of NMR signals from overlapping nanoscale regions of sample on the diamond surface. Only NV centres of the same orientation (shown in red), aligned along an externally applied static magnetic field, $B_0$, contribute to the ensemble NV NMR signal. {\bf d}, Larmor precessing nuclear spins in the sample produce an effective AC magnetic field (such as the one shown by the green line) that is detected by NV sensors in a frequency-selective manner using an XY8-$k$ pulse sequence.}
\end{figure*}

For both sensor modalities, an NV NMR measurement proceeds in the following way. First, an 8 $\mu$s long 532 nm laser pulse optically pumps the NV electronic spins into the $|0 \rangle$ state. Resonant microwave pulses are then applied to the NV electronic spins: first, a $\pi/2$-pulse prepares a coherent superposition of ground spin states $(|0 \rangle + |1 \rangle)/\sqrt{2}$; next, an XY8-$k$ sequence allows the NV spins to probe the local magnetic environment \cite{Gullion1}; and finally, a $\pi/2$-pulse projects the evolved NV spin coherence onto a $|0 \rangle$, $|1 \rangle$ state population difference, which is detected via the NV spin-state dependent fluorescence intensity after a 500 ns 532 nm laser pulse. The XY8-$k$ pulse sequence consists of a block of eight sequential $\pi$-rotation pulses repeated $k$ times (Fig.\ 1d) and serves two purposes. First, the sequence dynamically decouples NV spins from the background magnetic environment (e.g., spin impurities in diamond and other sources of magnetic noise), so that the NV spin coherence time $T_2$ is extended beyond the inhomogeneous dephasing time $T_2^{*}$ and the single Hahn-echo coherence time \cite{deLange1,Naydenov1,Ryan1,BarGill1,Cywinski1}. Second, the XY8-$k$ sequence gives the NV spins narrow-band sensitivity to NMR signals centered at frequency $\nu = 1/2\tau$, where $\tau$ is the delay time between $\pi$-pulses, and with detection bandwidth $\Delta \nu = 0.111/k\tau$ \cite{Bylander1,BarGill2,Hall1}. The presence of an NMR signal resonant with the XY8-$k$ sequence is detected as a spectrally-specific change in the NV optical fluorescence signal (see Methods for details).

Importantly, the strength of the NV NMR signal and the number of nuclear spins detected per NV are sensitively dependent on the NV depth and the density of nuclear spins in the sample. To calibrate NV depth, we used NV NMR measurements from protons in immersion oil (a well-understood sample with uniform $^1$H density) placed on the diamond surface, together with a model of magnetic field fluctuations at each NV centre induced by the ensemble of statistically polarized nuclear spins in the sample. We used the resulting NV depth and the nuclear magnetic field model to determine the number of sensed nuclear spins in each sample studied. Details of the NV depth calibration are given in the Methods and in Ref.\ 31.

\begin{center}
{\bf \large{Multi-Species NMR with Single NV}}\\
\end{center}

\begin{figure*}[t!]
\includegraphics[width=6.5in]{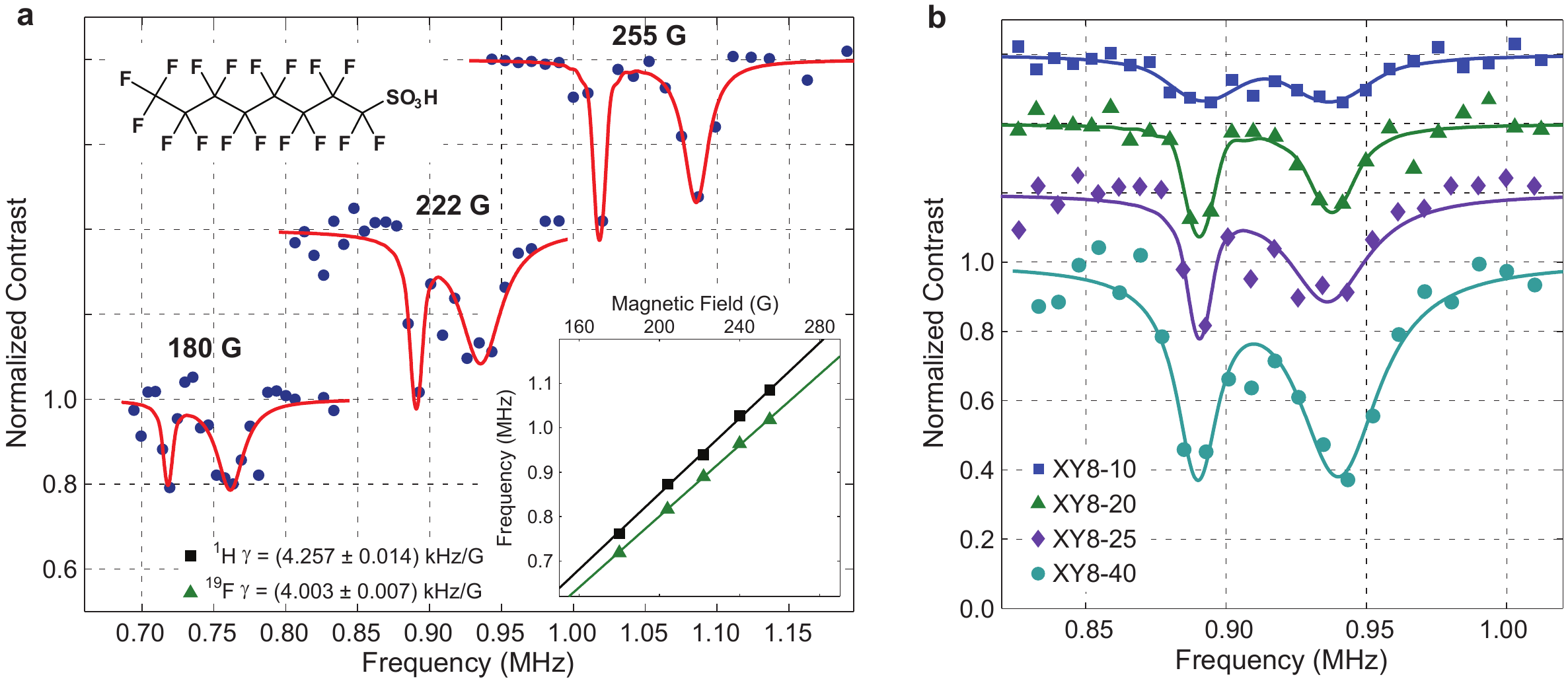}  
\caption{{\bf Multi-species nanoscale NMR with a single shallow NV centre.} {\bf a}, $^1$H and $^{19}$F NMR spectra of a fluorinated sample (PFOS/POSF) at several magnetic fields, measured with an XY8-10 sequence and fit with a model for the NV NMR lineshape. The PFOS molecular structure is indicated schematically. Inset: Measured $^1$H and $^{19}$F NMR resonance frequencies as a function of applied static magnetic field $B_0$. Linear fits yield gyromagnetic ratios in good agreement with the literature. {\bf b}, Series of NV NMR spectra for the fluorinated sample acquired with an increasing number of repetitions $k$ of the XY8-$k$ pulse sequence. Measurement sensitivity and spectral selectivity improve with increased repetitions $k$. (Note: in both {\bf a} and {\bf b}, spectra are offset vertically for clarity of display.)}
\end{figure*}
Fig.\ 2 shows example results for multi-species nanoscale NMR spectroscopy using a single NV centre, determined by the aforementioned calibration process to be $\sim$ 8 nm below the diamond surface. We placed a mixture of sodium perfluorooctanesulfonate (PFOS) and perfluorooctanesulfonyl fluoride (POSF) feedstock on the diamond surface and allowed it to dry under ambient conditions. Employing an XY8-10 pulse sequence, we then measured NV NMR spectra of the fluorinated residue and observed resonances corresponding to $^{19}$F and $^{1}$H nuclei over a range of applied static magnetic fields, $B_0$, oriented along the NV axis. Several representative NMR spectra are shown in Fig.\ 2a, where we fit the measured NV fluorescence to a model function in order to extract the frequencies and linewidths of the NMR resonance dips (see Methods and Ref.\ 31 for details). For the $^{19}$F NMR resonance, we determine that 50\% of the observed NV NMR signal results from $\sim$ 20,000 unpolarized fluorine nuclei in a $\sim$(10 nm)$^3$ volume, which has a statistical spin polarization equivalent to $\sim$ 140 polarized fluorine nuclei. In Fig.\ 2a we also plot the measured resonance frequency $\nu_0$ of each nuclear species as a function of $B_0$, with an observed linear dependence $\nu_0 = \frac{\gamma_n}{2 \pi} B_0$ that is consistent with the known gyromagnetic ratios of $^{19}$F and $^{1}$H \cite{Fuller1}.

To characterize the inhomogeneous dephasing time $T_2^{*}$ for each nuclear species, we varied the number of repetitions $k$ in the XY8-$k$ pulse sequence and observed the effect on the measured NV NMR resonance features. As shown in Fig.\ 2b, we found that increasing $k$, and thereby creating a narrower spectral filter for the NV NMR measurement, results in a narrowing and deepening of the resonance dips, setting lower limits  of $T_2^{*} \geq 32 $ $\mu$s for $^{19}$F and $T_2^{*} \geq 11$ $\mu$s for $^{1}$H for this nanoscale sample on the diamond surface.

\begin{center}
{\bf \large{Multi-Species NMR and MRI with NV Ensemble}}\\
\end{center}

As shown in Fig.\ 3, we acquired consistent multi-species nanoscale NMR spectra using an ensemble of high-density, shallow-implanted NV centres in a wide-field microscope setup, with the NV fluorescence signal detected by a CCD camera and integrated across the few-micron-wide laser spot. For this diamond chip, the mean lateral distance between NV centres of the same orientation is $\sim$ 30 nm (determined from the NV fabrication process and wide-field fluorescence measurements); and the mean NV depth is $\sim$ 10 nm (determined by the calibration process outlined above). In particular, the results in Fig.\ 3 demonstrate that high sensitivity nuclear spin sensing can be provided by NV ensembles, without choosing an optimal single NV sensor. In the first example (Fig.\ 3a), we again measured a fluorinated sample (PFOS/POSF) dried on the diamond surface. Both $^{1}$H and $^{19}$F NMR signals are resolved, albeit with broader linewidths than observed for the single NV centre data of Fig.\ 2. In the second example (Fig.\ 3b), we performed NV NMR measurements of $^{31}$P nuclei for a sample of powdered adenosine triphosphate disodium (ATP) salt on the diamond surface. The observed $^{31}$P NMR signal is weaker than for the $^{1}$H and $^{19}$F NV NMR measurements, due to the relatively smaller $^{31}$P gyromagnetic ratio and lower phosphorus spin density in the ATP salt sample. A plot of the measured $^1$H, $^{19}$F, and $^{31}$P resonance frequencies versus magnetic field (Fig.\ 3c) are in reasonable agreement with the known gyromagnetic ratios of these nuclear species \cite{Fuller1}. Note, however, that the effective gyromagnetic ratio for $^{31}$P derived from these NV NMR measurements is about 4\% higher than the free $^{31}$P value, which may result from dipole-dipole couplings within the ATP molecule.

\begin{figure*}[t!]
\includegraphics[width=6.5in]{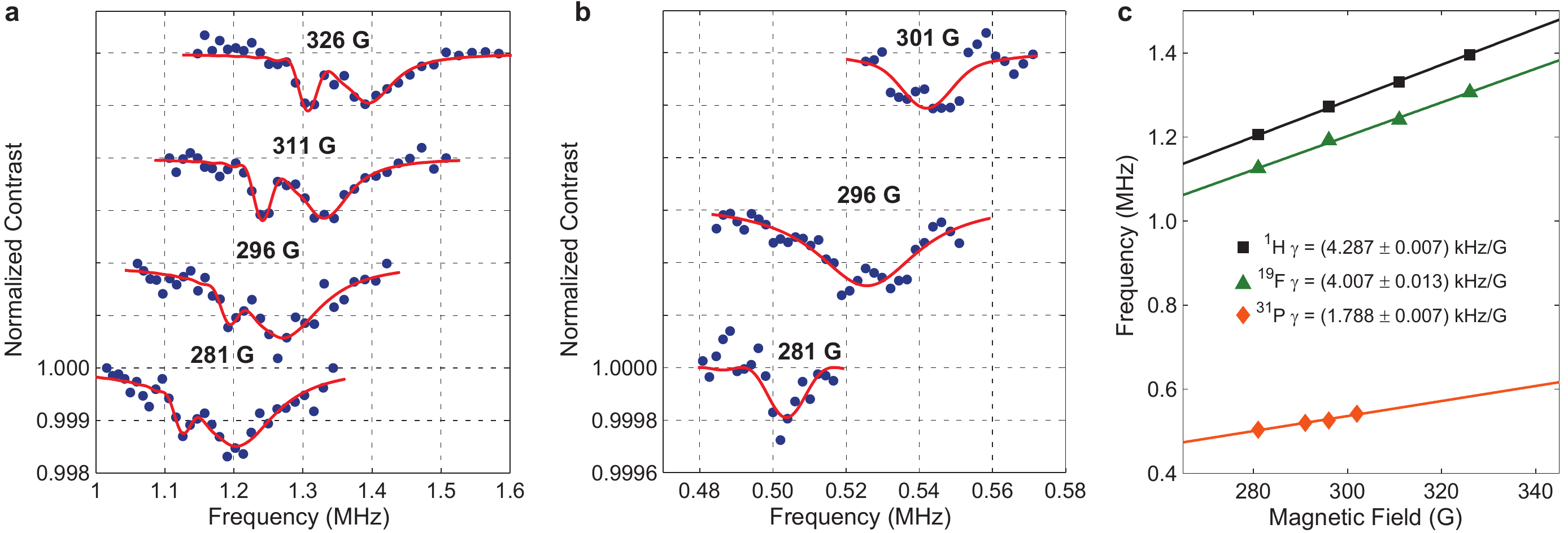}
\caption{{\bf Multi-species nanoscale NMR with a shallow NV ensemble.} {\bf a}, NMR spectra of a fluorinated sample (PFOS/POSF) at several magnetic fields. As with single-NV measurements, both $^{1}$H and $^{19}$F NMR signals are observed and fit with a model lineshape. {\bf b}, $^{31}$P NMR spectra from a powdered adenosine triphosphate (ATP) sample at several magnetic fields, smoothed and fit with the model lineshape. (Note: in both {\bf a} and {\bf b}, spectra are offset vertically for clarity of display.) {\bf c}, Measured $^1$H, $^{19}$F, and $^{31}$P NMR resonance frequencies as a function of applied static magnetic field $B_0$. Linear fits yield gyromagnetic ratios for $^1$H and $^{19}$F in good agreement with literature, with the value for $^{31}$P exceeding the literature value by about 4\%. (See discussion in main text.)}
\end{figure*}

We next used the wide-field NV microscope to demonstrate two-dimensional optical MRI of spatially-varying concentrations of $^{19}$F nuclear spins, again using the fluorinated sample (PFOS/POSF). We fabricated a patterned structure (mask) of SiO$_2$ on the diamond surface via atomic-layer deposition. This structure covered part of the diamond surface, with a sub-micron edge going from the full thickness of the SiO$_2$ layer (90 nm) to bare diamond. Fig.\ 4a shows a white light image of a corner defined by this structure. We introduced the fluorinated sample onto the diamond surface and applied the sensing protocol described above, which provided an NMR spectrum for each pixel of the CCD camera: i.e., optical MRI with about 500 nm lateral resolution, 50 $\mu$m field-of-view, and sensitivity to nuclear spins within $\sim 20$ nm of the diamond surface. An example $^{19}$F NMR image is shown in Fig.\ 4b; with single-pixel NMR spectra on the bare diamond surface and under the SiO$_2$  structure shown in Fig.\ 4c. See Methods for details. The SiO$_2$ structure prevented underlying NV centres from detecting the NMR signal from $^{19}$F nuclear spins in the sample, due to the strong ($1/d^3$) distance dependence of NV sensitivity to magnetic dipole fields. In contrast, the NV centres remained sensitive to $^{19}$F nuclear spins in the sample on the bare diamond surface. Note that a $^1$H NMR signal was observed across the full diamond surface even under the SiO$_2$ structure, consistent with other recent observations using NV diamond\cite{Rugar1, Wrachtrup1, Degen1, Mamin2} and MRFM\cite{degen2009nanoscale, mamin2009isotope, xue2011measurement}. These results illustrate the ability of our technique to provide nuclear-species-specific spectroscopic and imaging information for nanoscale samples across a wide field-of-view.

\begin{figure*}[t!]
\includegraphics[width=6.5in]{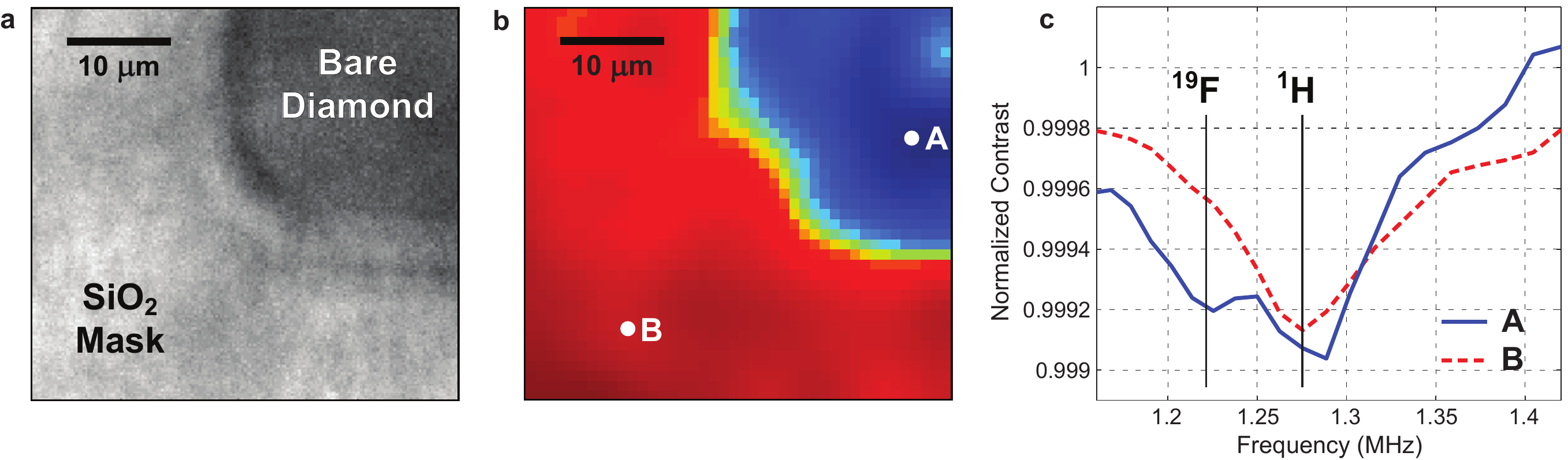}  
\caption{{\bf Optical MRI of multi-species sample with sub-micron structure.} {\bf a}, White-light transmission image of shaped SiO$_2$ structure (90 nm thick mask) on the surface of a diamond containing a shallow, high-density NV layer. {\bf b}, Optical MRI of $^{19}$F nuclear spin density in the fluorinated sample (PFOS/POSF) within $\sim$20 nm of surface. Blue represents a deep $^{19}$F NMR contrast dip and hence high fluorine concentration on the bare diamond surface as measured by the NV ensemble. Red represents no $^{19}$F NMR signal detected by the NV ensemble underneath the SiO$_2$ layer. {\bf c}, NV NMR spectra from two points of the image in {\bf b}. On the bare diamond surface (A), NMR signals are observed for both $^1$H and $^{19}$F. Under the SiO$_2$ structure (B), only ubiquitous surface layer $^1$H spins are detected, as the SiO$_2$ layer displaces the fluorinated sample $\sim$90 nm away from the diamond surface and NV sensors.}
\end{figure*}
\newpage
\begin{center}
{\bf \large{Probing Proton Layer on Diamond Surface}}\\
\end{center}

We next used a single NV confocal microscope to investigate the origin of the ubiquitous $^1$H NMR signal, observed on all diamond samples after extended exposure to air, including single NV and NV ensemble measurements, as well as in the presence of the dried PFOS/POSF sample and the SiO$_2$-coated region of the diamond. We applied Fomblin Y HVAC 140/13 oil, which contains ~40 $^{19}$F nuclei/nm$^3$ and no $^1$H, to the surface of a diamond directly after acid cleaning (see Methods). NV NMR measurements of the Fomblin oil yield strong NMR signals of both $^{19}$F and $^1$H nuclei for each of several NV centres probed (see example data in Fig.\ 5a). The different relative strengths of the $^{19}$F and $^1$H signals, dependent on the depth of the probed NV center (previously calibrated as described in the Methods and Ref.\ 31), are consistent with a thin adsorbed hydrocarbon or water layer on the diamond surface and below the thick layer of Fomblin oil (Fig.\ 5b,c). Applying the NV NMR lineshape model to this hypothesized sample geometry yields a proton-containing layer thickness of $0.8 \pm 0.2$ nm. The data are not consistent with an isotropic mixture of Fomblin and proton-containing molecules, which would result in $^1$H and $^{19}$F spectral signals having the same relative strengths for both NV centres. This experiment and analysis represents a form of nanoscale one-dimensional MRI, and it provides the first proof-of-principle demonstration of the capability of the NV NMR technique to extract thickness information for multi-layered thin films containing multiple nuclear spin species, with sub-nanometer resolution.

\begin{figure*}[t!]
\includegraphics[width=6.5in]{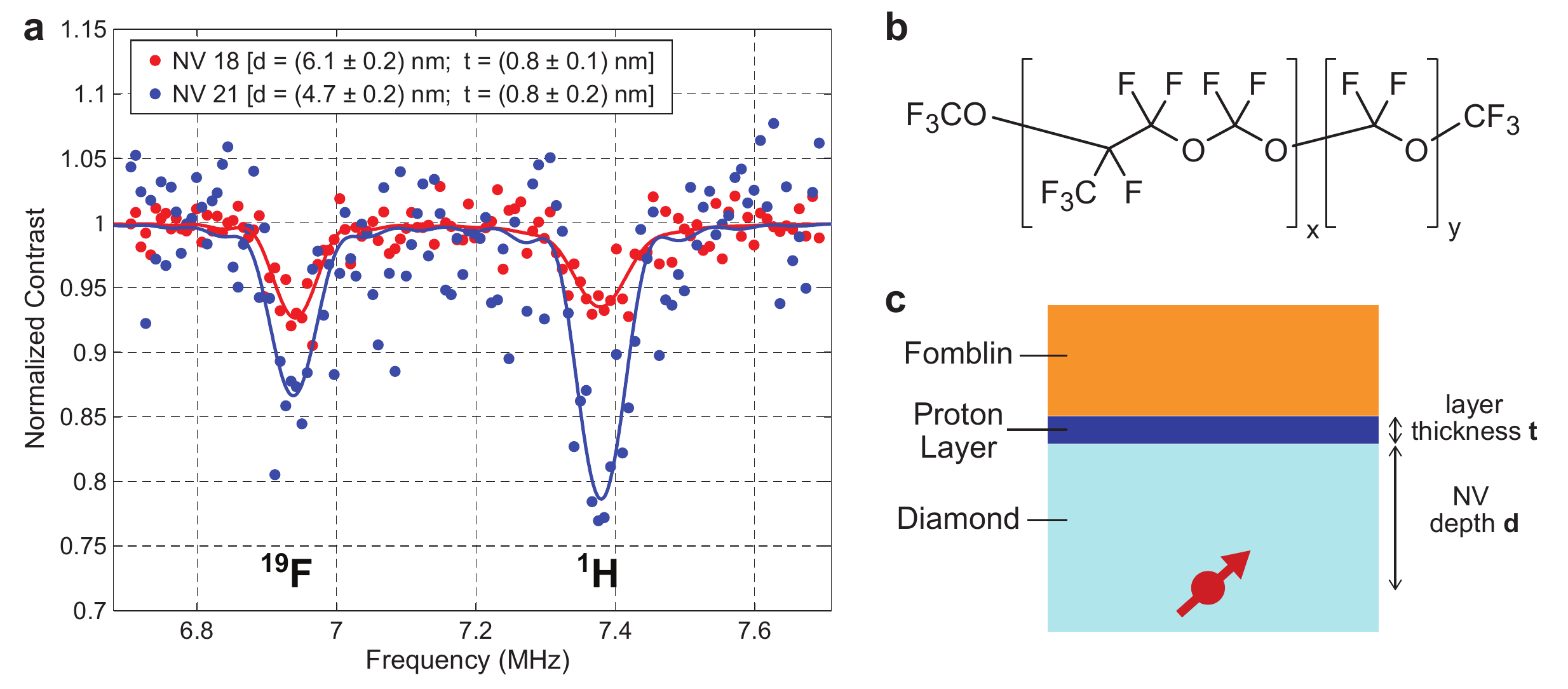}  
\caption{{\bf Determination of surface proton layer thickness.} {\bf a}, NMR signal from Fomblin Y oil on the diamond surface measured with two NV centres. NV depths are presented, as calculated from the lineshape model (see Methods). {\bf b}, Chemical structure of Fomblin Y oil. {\bf c}, Description of the hypothesized sample structure resulting from Fomblin Y oil floating above a thin layer of adsorbed hydrocarbons or water on the diamond surface.}
\end{figure*}
\newpage
\begin{center}
{\bf \large{Outlook}}\\
\end{center}

We demonstrated a new capability for nanoscale, optically-detected NMR spectroscopy and MRI of multiple nuclear species ($^1$H, $^{19}$F, $^{31}$P) using shallow NV centres in diamond. We performed simultaneous multi-species NMR spectroscopy under ambient conditions, employing two experimental modalities: (i) a scanning confocal microscope interrogating single NV centres, which is suitable for probing few-nanometre sized samples containing $\sim$ 100 polarized nuclear spins as well as extracting thickness information for multiple layers of thin films with sub-nanometer resolution; and (ii) a wide-field microscope using a CCD camera to image fluorescence from a high-density NV ensemble in a thin layer near the diamond surface, which is optimal for NMR spectroscopy and imaging over $>10$ $\mu$m field-of-view and with sub-micron resolution. Importantly, the NV ensemble results show that high-sensitivity nanoscale NMR does not require choosing an optimal single NV sensor. These complementary NV sensor modalities provide utility well beyond current NMR and MRI technology, opening the door to wide-ranging applications at the nanoscale, from studies of surface catalyst reactions to the identification of single protein structure and dynamics to functional MRI within living cells. Future challenges include improving the sensitivity and resolution of NV NMR and MRI, e.g., by realizing very shallow NV centres with good optical and spin properties and employing Fourier k-space imaging techniques with pulsed magnetic field gradients, respectively.

\medskip
\begin{center}
{\bf \large{Methods}}\\
\end{center}

\noindent {\bf Diamond samples.} The diamond for single NV measurements of PFOS/POSF was a $2 \times 2 \times 0.4$ mm 99.999\% $^{12}$C high-purity chemical vapor deposition (CVD) chip from Element 6 with an unpolished surface, implanted with 2.5-keV $^{14}$N$^{+}$ ions and annealed at 900 $^{\circ}$C for 8 hours. The NV centre used for the NV NMR measurements shown in Fig.\ 2 was in a region with a 2D NV density of $8 \times 10^{7}$ cm$^{-2}$, and its Hahn-echo $T_2$ was 26 $\mu$s. 
The diamond for single NV measurements of Fomblin was a $4 \times 4 \times 0.5$ mm 99.999\% $^{12}$C high-purity chemical vapor deposition (CVD) chip from Element 6 with an unpolished surface, implanted with 2-keV $^{15}$N$^{+}$ ions at a dose of $1 \times 10^{9}$ cm$^{-2}$. It was annealed at 800 $^{\circ}$C for 8 hours and cleaned in a three-acid mixture (1:1:1 nitric:sulfuric:perchloric acids).
The diamond for NV ensemble measurements was a $4 \times 4 \times 0.3$ mm 99.6\% $^{12}$C CVD chip from Element 6 implanted with 6 keV $^{14}$N$^{+}$ ions at a dose of $2 \times 10^{13}$ cm$^{-2}$. The diamond was annealed at 800 $^{\circ}$C for 2 hours, producing NV centres with a 2D density of $3.5 \times 10^{11}$ cm$^{-2}$ in a $\sim$ 10 nm thick layer at an average depth of $\sim$ 10 nm, estimated by Monte Carlo simulations. The ensemble Hahn-echo $T_2$ was 3 $\mu$s.

\noindent {\bf SiO$_2$ structure.} The diamond used for NV ensemble NMR imaging (Fig.\ 4) was prepared by cleaning in piranha solution (2 parts H$_2$SO$_4$ to 1 part H$_2$O$_2$ v/v) for more than 1 hour. Using atomic layer deposition (ALD, Savannah Atomic Layer Deposition S200), a 3 nm layer of Al$_2$O$_3$ was grown on the diamond surface followed by a 90 nm layer of SiO$_2$. The deposition temperature of the substrate was 250 $^{\circ}$C, and deposition rate was 0.5 nm/min. The SiO$_2$-coated diamond was cleaned with acetone and isopropanol and then baked for 2 minutes on a hot plate at 115 $^{\circ}$C to remove water. The diamond was then spin coated with hexamethyldisilazane followed by photoresist S1805 and once again heated at 115 $^{\circ}$C for 90 seconds. It was exposed with a photomask using a Suss MicroTec MJB4 and developed with CD26 for 45 s, rinsed with deionized water, and blown dry with nitrogen. SiO$_2$ was etched with buffered oxide etchant (BOE) for 1 minute at an etch rate of 330 nm/min. Photoresist was then removed by soaking in acetone for 5 minutes. Finally, the diamond was cleaned in piranha solution for 1 hour.

\noindent {\bf Fluorine samples.} The PFOS sample was prepared by mixing 2 mmol of solid sodium hydroxide with 2 mmol of liquid perfluorooctanesulfonyl fluoride (POSF) (Sigma Aldrich, St.\ Louis, MO) to produce a mixture containing sodium perfluorooctanesulfonate (PFOS). The resulting mixture was applied to the diamond surface and allowed to dry, leaving a solid sample of PFOS on the surface. For measurements of the protonated surface layer, Fomblin Y HVAC 140/13 oil was applied directly to the diamond surface. 

\noindent {\bf Confocal microscope.} Measurements of PFOS/POSF with single NV centres (Fig.\ 2) were performed using a custom-built scanning confocal microscope. Optical excitation was provided by an 800 mW 532 nm diode pumped solid-state (DPSS) laser (Changchun New Industries Optoelectronics Tech MLLIII532-800-1) focused onto the diamond using a $100\times$, 1.3 NA oil immersion objective (Nikon CFI Plan Fluor $100\times$ oil). The laser power incident on the sample was 0.5 mW. The excitation laser was pulsed by focusing it through an acousto-optical modulator (Isomet 1205C-2). NV fluorescence was collected through the same objective and separated from the excitation beam using a dichroic filter (Semrock LM01-552-25). The light was additionally filtered (Semrock LP02-633RS-25) and focused onto a single-photon counting module (Perkin-Elmer SPCM-ARQH-12). Microwaves were delivered to the diamond using a 900 $\mu$m-diameter loop fabricated on a glass cover slip, with the diamond glued to the cover slip and in contact with the loop. The loop was driven by an amplified (Mini-circuits ZHL-16W-43-S+) microwave synthesizer (Windfreak SynthNV). The phase of microwave pulses was controlled using an in-phase/quadrature (IQ) mixer (Marki IQ1545LMP). Microwave and optical pulses were controlled using a computer-based digital delay generator (SpinCore PulseBlaster ESR400). Measurement protocols (pulse sequences, data acquisition, etc.) were controlled by custom software. The static magnetic field was applied with a permanent magnet whose distance and position relative to the NV centre was controlled with a three-axis stage.

Measurements of Fomblin with single NV centres (Fig.\ 5) were performed with a similar confocal microscope in which microwave pulses were generated with a Tektronix AFG3052C arbitrary waveform generator. Phase was controlled with an IQ mixer driven by a second Tektronix AFG3052C AWG. Microwaves were delivered via a stripline fabricated on a glass coverslip and placed against the diamond surface.

\noindent {\bf Wide-field microscope.} Measurements with NV ensembles were performed using a custom-built wide-field microscope. Optical excitation was provided by a 3W 532 nm LaserQuantum mpc6000 laser focused through a glass coverslip and the diamond chip onto the opposite diamond surface, containing the shallow, high-density NV layer, by a 100$\times$ 0.9 NA air objective (Olympus MPlan N). The laser power incident on the sample was 800 mW. The diamond was attached to the coverslip with Norland Blocking Adhesive 107 (Norland, Cranbury, NJ), which was cured under a UV lamp for 30 minutes. The laser was controlled with an acousto-optical modulator (Isomet M1133-aQ80L-1.5). The NV fluorescence signal was collected through the same objective and separated from the excitation beam with a dichroic mirror (Semrock LM01-552-25) and optical filters (Semrock LP02-633RS-25 and FF01-750SP-25) before being imaged onto a CCD camera (Starlight Express SXVR-H9). An optical chopper was used to block fluorescence during optical state preparation of the NV centres. Microwaves were synthesized with a signal generator (Agilent E8257D), amplified (Mini-circuits ZHL-16W-43-S+), and applied to the sample with a small wire loop placed against the diamond. The microwave pulse phase was controlled by an IQ mixer (Marki IQ1545LMP). Microwave and optical pulses were controlled by a pulse generator (SpinCore PulseBlasterESR-PRO 500 MHz) governed by custom software. The static magnetic field was applied with a permanent magnet whose distance and position relative to the NV ensemble was controlled with a three-axis stage.

The laser spot on the diamond had a FWHM size of $\sim$ 60 $\mu$m. For the white light image, each CCD pixel represented 200 nm x 200 nm on the diamond surface. For the NV fluorescence measurement, each CCD pixel represented 1 x 1 $\mu$m on the diamond surface, although the point spread function of the detection optics was $\sim$ 500 nm. Smaller CCD pixels could be used, with reduced SNR. Each NV NMR measurement average was performed for 500 ms (2000 chopper cycles) at each dynamical decoupling delay, and a full dataset consisted of $\sim$ 800 averages. For simple spectroscopic measurements, a 26 $\mu$m x 20 $\mu$m field of view was sampled and the measurements from each pixel were averaged together before further processing. For imaging, a larger field of view was sampled and each pixel was analyzed separately. For display, as in Fig.\ 4b, the image was further processed with a 3-pixel width Gaussian blur.

\noindent {\bf NV NMR measurements, spectral model, and NV depth estimation.} Spin-state measurements take place in the NV ground electronic state, in which the $|0\rangle$ spin state is split from $|1\rangle$ and $|-1\rangle$ spin states by a zero-field splitting of 2.87 GHz, and $|1\rangle$ and $|-1\rangle$ experience Zeeman splitting in the presence of an external magnetic field. NV fluorescence is induced with a 532 nm laser pulse, with a stronger signal when spins are in state $|0\rangle$, as well as optical pumping into $|0\rangle$, due to non-radiative decay from the $|1\rangle$ and $|-1\rangle$ excited electronic states through metastable singlet states and then into $|0\rangle$.

For the NV NMR experiments described here, the magnetic signal of interest is produced by nuclear spins on the diamond surface interacting with shallow NV centres through magnetic dipole-dipole coupling. The specific components of the dipole-dipole Hamiltonian that are responsible for the measured signal stem from the $S_z I_x$ and $S_z I_y$ terms, where $S_z$ is the $z$ component of the NV spin (defined by the NV symmetry axis) and $I_{x,y}$ are the $x$ and $y$ components of the nuclear spin. These terms couple the NV spin to the transverse component of the nuclear spin, which precesses in a static magnetic field at the nuclear Larmor frequency. A nearby permanent magnet aligned with the NV centre quantization axis sets the static magnetic field, $B_0$. A single measurement consists of repeating the optical pumping, XY8-$k$ sequence, and optical detection a few hundred times in order to collect sufficient photons at the detector. The measurement is then repeated for a series of XY8-$k$ pulse delay times, $\tau$, to determine the spectrum of the magnetic environment. When $\tau$ matches a half-period of the nuclear spin precession, the magnetic coupling effectively drives the NV spin away from the initial spin state, which is detected as a change in NV fluorescence intensity. These dips in the signal occur at the Larmor frequencies of nuclei on the surface (or other sources of noise, such as nuclear impurities within the diamond).

To perform NV NMR measurements, two fluorescence measurements $F_1$ and $F_2$ were acquired for each pulse sequence delay with the final $\pi/2$ pulse 180$^{\circ}$ out of phase. This procedure removes common-mode noise from laser intensity fluctuations. Normalized contrast, $C$, was then calculated as
\begin{equation}
C = \frac{F_2 - F_1}{F_2 + F_1}.
\end{equation}
The broad decrease in contrast resulting from intrinsic NV decoherence was removed with a linear baseline correction. The corrected contrast was fit with the function
\begin{equation}
C(\omega) = \exp \left(-\sum\limits_{i} \chi_i(\omega) \right),
\end{equation}
where $\chi_i(\omega)$ describes the NV decoherence due to each nuclear species, $i$. It is a function of the frequency-dependent variance in the magnetic field signal (spectral density), $\langle \left| B_{z}^i(\Omega,\omega_L) \right|^2 \rangle$, created by the nuclear spins, as well as a function $g(\Omega,\tau,N)$ describing the NV sensor response to the pulse sequence:
\begin{equation}
\chi_i(\omega) = \frac{\gamma_e^2}{4\pi}  \int_{-\infty}^{+\infty} \langle \left| B_{z}^i(\Omega,\omega_L) \right|^2 \rangle \left| g(\Omega,\tau,N) \right|^2 \mathrm{d} \Omega.
\end{equation}
The function $g(\Omega,\tau,N)$ is the Fourier transform of $g(t)$, where $g(t)$ is a function describing the sign of NV spin phase accumulation during the pulse sequence. For the primary resonance of the XY8-$k$ sequence,
\begin{equation}
 \left| g(\Omega,\tau,N) \right|^2 \approx \frac{4}{\pi^2}(N \tau)^2 \text{sinc}^2 \left(\frac{N \tau}{2} \left(\Omega - \frac{\pi}{\tau} \right)  \right).
\end{equation}
The magnetic signal created by a semi-infinite layer of spin-1/2 nuclei with density $\rho$ near an NV centre oriented along the [1 0 0] crystallographic axis is
\begin{equation}
\langle \left| B_{z}(\Omega,\omega_L) \right|^2 \rangle = \rho \frac{5 \pi}{48}\left( \frac{\mu_0 \hbar \gamma_n}{4 \pi} \right)^2 \left(\frac{1}{(d_{NV}+z_1)^3}-\frac{1}{(d_{NV}+z_2)^3}\right) \frac{T_2^{*-1}}{(\Omega - \omega_L)^2 + (T_2^{*-1})^2},
\end{equation}
where $\omega_L$ is the nuclear Larmor frequency, $T_2^{*}$ is the nuclear spin dephasing time, $d_{NV}$ is the depth of the NV centre below the diamond surface, $z_1$ is the distance from the diamond surface to the lower bound of the layer, and $z_2$ is the distance from the diamond surface to the upper bound of the layer. Combining these expressions and using the relationship $\omega = \pi/\tau$ for the filter resonance condition gives
\begin{equation}
\chi_i(\omega) = \rho_i \frac{5}{48 \pi} \left(\frac{\mu_0 \gamma_{n,i} \gamma_e \hbar}{4\pi} \right)^2 \left(\frac{1}{(d_{NV}+z_1)^3}-\frac{1}{(d_{NV}+z_2)^3}\right) I_i(\omega),
\end{equation}
where $I_i(\omega)$ is the convolution between the Lorentzian lineshape of the nuclear spin signal from species $i$ and the sinc$^2(\omega)$ lineshape of the filter function for the XY8-$k$ sequence. It can be expressed as
\begin{widetext}
\begin{multline}
I_i(\omega) = \frac{2 T_{2,i}^{*2}}{\left[1 + 
   T_{2,i}^{*2} \left(\omega_{L,i} - \omega\right)^2\right]^2} \left\{e^{-\frac{N \pi}{
    \omega T_{2,i}^{*}}} \left[\left[1 - T_{2,i}^{*2} \left(\omega_{L,i} - \omega\right)^2\right] \cos \left[
        \frac{N \pi}{\omega} \left(\omega_{L,i} - \omega\right)\right]\right.\right.\\ -
      \left.\left. 2 T_{2,i}^{*} \left(\omega_{L,i} - \omega\right) \sin \left[
        \frac{N \pi}{\omega} \left(\omega_{L,i} - \omega\right)\right]\right] -1 + 
     \frac{N \pi}{\omega T_{2,i}^{*}} \left[1 + T_{2,i}^{*2} \left(\omega_{L,i} - \omega\right)^2\right] +
      T_{2,i}^{*2} \left(\omega_{L,i} - \omega\right)^2 \right\},
\end{multline}
\end{widetext}
where $ N = 8 k$ is the total number of $\pi$-pulses. For additional details see ref.\ 31.

To estimate the depth of single NV centres, NV NMR measurements were performed with a drop of immersion oil (Olympus Type-F Low Auto-Fluorescence) on the diamond surface. The parameters $d_{NV}$ and $T_2^{*}$ were determined by fitting the NV NMR signal contrast with equation (2) using a density\cite{Degen1} $\rho = 60$ protons/nm$^3$ and assuming $z_1 = 0$ and $z_2 \rightarrow \infty$.

To calculate the thickness of the proton layer described in Fig.\ 5, we simultaneously fit the $^1$H and $^{19}$F spectral dips using equation (2). We assumed a $^1$H layer of finite thickness with density $\rho = 60$ protons/nm$^3$ between the diamond and a semi-infinite layer of Fomblin Y oil with density $\rho = 40$ fluorines/nm$^3$. The proton layer thickness, $t$, and NV depth, $d$, were left as free parameters determined by fitting. Here, we assumed that the noise spectrum of both the Fomblin oil and the proton layer were given by delta-functions in frequency space. By observing the scaling of the NMR signals with the number of applied pulses, we determined that both nuclear dephasing timescales were much longer than those probed in this experiment. We then tested an alternative model in which both $^1$H and $^{19}$F were isotropically distributed in a semi-infinite layer on the diamond surface. We found that for NV centres of different depths, different ratios of $^1$H:$^{19}$F nuclear spins were needed to achieve a fit. This was inconsistent with the fact that the NVs were interrogating the same sample, and so this hypothesis was rejected. 

\noindent {\bf Acknowledgments} This work was supported by the NSF and the DARPA QuASAR programme. F.C. acknowledges support from the Swiss National Science Foundation (SNSF). I.L. acknowledges support from a NDSEG fellowship.

\noindent {\bf Author Contributions} S. J. DeVience and L. M. Pham contributed equally to this work. R.L.W., S.J.D., L.M.P., and N.B.-G. conceived the idea of the study. S.J.D., L.M.P., I.L., A.O.S, and M.C. performed the measurements and analyzed the data. F.C. and S.J.D. developed the model for describing the signal. H.Z. and C.B. created the SiO$_2$ masks. M.D.L., H.P., R.L.W. and A.Y. conceived the NV-diamond wide-field magnetic imager and its applications. All authors discussed the results and participated in writing the manuscript.

\noindent {\bf Competing Financial Interests} The authors declare no competing financial interests.


\end{document}